\documentclass[manuscript]{aastex}

\def\feka{Fe K$\alpha$}
\def\chandra{{\it Chandra}}
\def\xmm{{\it XMM-Newton}}
\def\asca{{\it ASCA}}

\def\rosat{{\it ROSAT}}

\def\integral{{\it INTEGRAL}}

\def\vlba{{\it VLBA}}
\def\glast{{\it GLAST}}
\def\swift{{\it Swift}}
\def\suzaku{{\it Suzaku}}

\def\lum{erg s$^{-1}$}
\def\flux{erg cm$^{-2}$ s$^{-1}$}
\def\nh{cm$^{-2}$}
\def\arcsec{$^{\prime\prime}$}
\def\deg{$^{\circ}$}

\def\ltsima{$\; \buildrel < \over \sim \;$}
\def\simlt{\lower.5ex\hbox{\ltsima}} 
\def\gtsima{$\; \buildrel > \over \sim \;$}
\def\simgt{\lower.5ex\hbox{\gtsima}} 

\input{psfig.tex}

\begin{document}
\title{Discovery of an Extreme MeV Blazar with the Swift Burst Alert Telescope} 


\author{R.M. Sambruna, C. B. Markwardt, R.F. Mushotzky, J. Tueller, R.
Hartman}
\affil{NASA's GSFC, Greenbelt, MD 20771}

\author{W.N. Brandt, D.P. Schneider, A. Falcone, A. Cucchiara} 
\affil{Department of Astronomy \& Astrophysics, The Pennsylvania State
University, University Park, PA 16802} 

\author{M. F. Aller, H. D. Aller} 
\affil{University of Michigan, Ann Arbor, MI 48109-1042} 

\author{I. Torniainen}
\affil{ Mets\"ahovi Radio
Observatory,  Helsinki University of Technology, Mets\"ahovintie 114,
FIN-02540 Kylm\"al\"a, Finland}

\author{F. Tavecchio, L. Maraschi}
\affil{Osservatorio Astronomico di Brera, via Brera 28, 20121 Milano,
Italy}

\author{M. Gliozzi}
\affil{George Mason University, Dept. of Physics \& Astronomy and School of
Computational Sciences, MS 3F3, 4400 University Drive, Fairfax, VA 22030}

\author{T. Takahashi} 
\affil{Institute of Space and Astronautical Science, 3-1-1 Yoshinodai,
Sagamihara-shi, Kanagawa 229-8510, Japan}

\begin{abstract}

The Burst Alert Telescope (BAT) onboard \swift\ detected bright
emission from 15--195~keV from the source SWIFT~J0746.3+2548 (J0746 in
the following), identified with the optically-faint (R $\sim$ 19),
$z$=2.979 quasar SDSS J074625.87+244901.2. Here we present \swift\ and
multiwavelength observations of this source. The X-ray emission from
J0746 is variable on timescales of hours to weeks in 0.5--8~keV and of
a few months in 15--195~keV, but there is no accompanying spectral
variability in the 0.5--8 keV band. There is a suggestion that the BAT
spectrum, initially very hard (photon index $\Gamma \sim 0.7$),
steepened to $\Gamma \sim 1.3$ in a few months, together with a
decrease of the 15--195~keV flux by a factor $\sim$ 2. The 0.5--8~keV
continuum is well described by a power law with $\Gamma \sim 1.3$, and
spectral flattening below 1~keV. The latter can be described with a
column density in excess of the Galactic value with intrinsic column
density N$_H^z \sim 10^{22}$ \nh, or with a flatter power law,
implying a sharp ($\Delta \Gamma$ \gtsima 1) break across 16~keV in
the quasar's rest-frame. The Spectral Energy Distribution of J0746 is
double-humped, with the first component peaking at IR wavelengths and
the second component at MeV energies. These properties suggest that
J0746 is a a blazar with high gamma-ray luminosity and low peak energy
(MeV) stretching the blazar sequence to an extreme. 

\end{abstract}

\keywords{Galaxies: active --- galaxies: 
quasars: individual (J0746.3+2548)--- X-rays: galaxies} 

\section{Introduction}

Among Active Galactic Nuclei (AGN), blazars are characterized by the
most extreme properties, such as rapid variability, large apparent
isotropic luminosity, high and variable polarization in the radio and
optical, and emission across the entire electromagnetic spectrum, from
radio wavelengths to GeV and TeV  gamma-ray energies (e.g., Ulrich,
Maraschi, \& Urry 1997). These properties are ascribed to non-thermal
(synchrotron and inverse Compton) emission from a relativistic jet 
seen nearly end-on. Blazars' Spectral Energy Distributions (SEDs) are
double-humped in a power-per-decade representation, with the
synchrotron component peaking anywhere from IR to X-rays, and the
inverse Compton component extending up to gamma-rays. The SEDs form a
sequence in luminosity, with more luminous sources having both peaks
at lower energies than their fainter counterparts (Fossati et
al. 1998, Ghisellini et al. 1998). 

Traditionally, blazars have been selected at radio, optical, and X-ray
wavelengths according to various criteria (e.g., Laurent-Mueheleisen
et al. 1997, Perlman et al. 1998). Only a few sources were discovered
using gamma-rays, both at GeV (Romani et al. 2004) and MeV wavelengths
(Blom et al. 1995). The advent of detectors with improving sensitivity
at hard X-ray energies ($>$ 10 keV), is now providing a new window for
blazar discovery and observation.
 
Such a window was recently opened by the Burst Alert Telescope (BAT;
Barthelmy et al. 2005) onboard the \swift\ satellite. On December 15,
2004, the BAT began an ongoing monitoring of the sky with its
wide-angle (2 steradians) field of view and high sensitivity in
the 15--200~keV energy band. During the first 3 months of the sky survey
operations, with a flux threshold of $\sim 10^{-11}$ \flux, the BAT
detected several AGN, both radio-loud and radio-quiet (Markwardt et
al. 2006). Here we describe the BAT discovery and subsequent \swift\
and multiwavelength observations of a new blazar with extreme
properties, SWIFT~J0746.3+2548 (J0746 in the following). The source
was identified with an optically-faint quasar at $z$=2.979 in the
Sloan Digital Sky Survey (SDSS) archive. In this paper, we will
present evidence that J0746 is a blazar (Flat Spectrum Radio Quasar)
and more specifically, a new member of the MeV blazar
class\footnote[1]{We define an MeV blazar as a blazar with peak of the
Compton component in the SED at MeV energies.}. 

In Fall 2005, J0746 was the target of an X-ray (\swift\ and
\suzaku), optical, and radio campaign. 
The radio observations were taken with the 26-m diameter U. of
Michigan Radio Astronomy Observatory (UMRAO) and the 14-m diameter
Mets\"ahovi radio telescopes, while optical-UV photometry is from the
\swift\ UVOT telescope.  Here we present the \swift\ and radio
observations, while a complete account of the
\suzaku\ data will be forthcoming (Takahashi et al. 2006, in prep.). We also
report on a new optical spectrum taken with the {\it Hobby-Eberly Telescope} (HET).

This paper is structured as follows. In \S~2 we describe the source
discovery and identification. The observations and data reduction
procedures are presented in \S~3, while in \S~4 we give the results,
in \S~5 the discussion, and in \S~6 the conclusions. We assume a
concordance cosmology with H$_0=71$ km s$^{-1}$ Mpc$^{-1}$,
$\Omega_{\Lambda}$=0.73, and $\Omega_m$=0.27 (Spergel et
al. 2003). The energy index $\alpha$ is defined such that $F_{\nu}
\propto \nu^{-\alpha}$. 
 
\section{Discovery of J0746 at hard X-rays} 

During the first 3 months of its sky survey, the BAT discovered hard
X-ray emission from several extragalactic sources (Markwardt et
al. 2006). With a flux F$_{20-200~keV} \sim 9 \times 10^{-11}$ \flux\
and a $>7\sigma$ detection (Fig. 1), SWIFT~J0746.3+2548 stands out as
one of the brightest. The follow-up observations with the XRT in
0.5--10 keV shows the presence of a source at RA(2000)= 07hrs 46m
25.6s, DEC(2000)=+25deg 49m 02.0s, or 1\arcmin\ from the BAT
position. The radii of the error circles for the BAT and XRT positions
are 2\arcmin\ and 3\arcsec, respectively. No other X-ray sources are
present in the XRT field of view, leading us to identify the XRT
source with the origin of the BAT flux.

A comparison with the latest Sloan Digital Sky Survey (SDSS; York et
al. 2000) Quasar Catalog (Schneider et al. 2005) yields a match with
SDSS~J074625.87+254902.2, located at RA(2000)=07h46m25.9,
DEC(2000)=+25d49m02.1s; the accuracy of the SDSS coordinates is
approximately 0.1\arcsec.  The SDSS counterpart of the X-ray source is
an optically faint (R $\sim$ 19 mag) quasar at a redshift
$z$=2.979. Using optical (Richards et al. 2006) and X-ray (Bauer et
al. 2003) number counts, we estimate that the false-match probability
of the XRT and SDSS sources is $P\sim 6.2
\times 10^{-6}$. Thus, the identification of the \swift\ source with
SDSS~J074625.87+254902.2 (J0746 in the following) is regarded as
robust.

Because of its optical faintness, J0746 largely escaped the attention
of the Astronomy community. A search of the \verb+NED+ database shows
23 references. Radio \vlba\ observations at 2.3 and 8.6 GHz show the
quasar to be core-dominated, with no obvious extended radio features
on mas scales (Fey \& Charlot 2003). From the data in this paper, we
calculate a radio-loudness parameter, $\log R_L
\sim 3$, in the rest-frame of the quasar. The radio-loudness parameter
is defined as the ratio of the 5~GHz to the $V$--band flux. To
K-correct the radio and optical fluxes we used power law continua with
$\alpha_{\nu, radio}=0.08$ (see \S~3.4) and $\alpha_{\nu,opt}=0.54$
(\S~3.3).

Although J0746 was only recently identified as a quasar, the archival
literature contains a number of observations of the object obtained
during wide-angle surveys at a number of wavelengths.  In the radio,
J0746 was monitored at 8 and 2.3~GHz with the Green Bank
Interferometer in the time period 1988--1995 (Lazio et al. 2001). No
variability was detected at 2~GHz. However, the light curve at 8~GHz
(32~GHz in the rest-frame) shows a slow but smooth increase of the
flux over 3 years, with a max/min flux change of a factor of 3. The
radio spectral index steepens on the same timescale (Fig. 8 of Lazio
et al. 2001).

The only previous X-ray observation of J0746 was by the \rosat\
satellite during its All-Sky Survey (RASS). Weak 0.2--2~keV emission,
with 16 counts in 438 sec, was recorded. Since the Galactic column
density in the direction of J0746 is low, N$_H^{Gal}=4.04
\times 10^{20}$ \nh, this implies a highly absorbed AGN;
alternatively, \rosat\ caught the quasar in a low intensity state. 

Based on its broad-band properties, we argue in this paper in favor of
a blazar nature of J0746 (\S~5). 

\section{Observations} 

\subsection{X-rays} 

The \swift\ observations of J0746 are summarized in Table~1, where the
BAT observations and follow-up XRT pointings are reported. The
November 4 XRT observation was performed contemporaneously with
\suzaku\ and radio measurements. 

\noindent{\bf BAT:} J0746 was in the
BAT field of view for a total of 755~ks since the beginning of the
survey on December 15, 2004. The results described here are for the
first 9 months of the survey, or until the end of September, 2005. The
average count rate from 15--195~keV is listed in Table~1.

The reduction of the BAT data was described in Marqwardt et
al. (2005).  Figure~1 shows the BAT significance map, ranging from 
$-2\sigma$ (white) to $+8\sigma$ (black). Pixel size is
5\arcmin. The cross marks the SDSS position of J0746. The source is
detected at $>7\sigma$.

We extracted two BAT spectra, one for the initial 3.5 months of the
survey, when J0746 was first discovered, and one integrated over the
whole 9 months. The BAT spectrum of J0746 is based on the mosaicked
sum of all observations which contain the source (mid Dec 2004 - mid
Sep 2005).  The minimum partial coding threshold was 15\%. Before
mosaicking, the images were corrected for systematic off-axis
attenuation effects, such that the spectrum should remain relatively
constant with off-axis angle (calibrated by BAT observations of the
Crab).

The BAT spectra were fitted with \verb+XSPEC+ v.11.3.2, with the
response matrix based on the latest (December 2005) calibration
products. Because of the spectral extraction protocol, the BAT
spectrum consists of 4 bins containing the source count rates in 4
energy bands: 14--25~keV, 25--50~keV, 50--100~keV, and 100--200~keV.

\noindent{\bf XRT:}
The XRT observed J0746 four times (Table~1). During the first three
observations in September, which overlapped with the BAT observations
(Fig.~2a), short exploratory exposures were used. On November 4, XRT
observed J0746 for 15~ks, together with the UVOT,
\suzaku, and ground-based radio observations. The XRT count rates in the
four epochs are listed in Table~1.

We used the event2 files provided by the \verb+HEASARC+, cleaned
according to standard criteria (Capalbi et al. 2005). Only data
acquired through Photon Counting (PC) mode, which has the maximum
sensitivity, were used; no pileup is present at the source's count
rates (Table~1). Light curves and spectra were extracted from a
circular region centered on the XRT position and with radius
20\arcsec. The background was extracted from a circle with radius
130\arcsec\ in a region free from obvious X-ray sources. Inspection of
the background light curve shows no variability within the four XRT
exposures.

The XRT spectra were rebinned in order to have at least 20 counts in
the new bins, except for the first observation on 2005 September 5
(OBSID 002) where the spectrum was rebinned with a minimum of 10
counts/bin, to increase the number of data points at lower energies
($<1$ keV) for a meaningful determination of the column density. The
spectra were fitted within \verb+XSPEC+ v.11.3.2 with the calibration
files provided by the \verb+HEASARC+, in the energy range 0.5--8 keV,
where the calibration is best known and the background negligible.

\subsection{Optical/UV photometry} 

\noindent{\bf UVOT:} 
The \swift\ UVOT observed the source continuously (except for Earth
occultations) starting at UT Nov 4 00:39 until UT Nov 6 15:59
(Table~2).  During each orbit data were collected from the six filters,
from the Optical to Ultraviolet wavelengths.  Since the source is
faint, we coadded all the data collected on the first day (Nov 4) in
only one image for each filter. Similarly, the Nov 5 and Nov 6 data
were coadded. The log of the UVOT observations, together with the
fluxes, is given in Table~2.

The data analysis was performed using the \verb+uvotsource+ task
included in the latest HEASOFT software release
(http://swift.gsfc.nasa.gov/docs/software/lheasoft/).  The background
was subtracted, and corrections for coincidence loss effects (similar
to pileup for the XRT) were applied. The magnitudes are converted into
fluxes using the latest in-flight flux calibration factors and
zero points.  The source is well detected in the optical filters with
over $3\sigma$ confidence level.  There is no detection in the UV
images, and only upper limits are reported in Table~2.

\noindent{\bf SDSS:} Table~2 presents optical photometric 
measurements of J0746 obtained from the SDSS DR3 catalog in the SDSS
system (Fukugita et al. 1996) on 19 December 2001. All fluxes are
dereddened for Galactic absorption. The SDSS fluxes are consistent
with those from the UVOT at similar wavelengths.

A search of the USNO archive for J0746 returns the following
historical records of its optical flux: $B$=20.01, 19.87 mag;
$R$=19.16, 19.04 mag; and I=18.92 mag. The $B$-band measurements are
consistent with those from the UVOT. 

\subsection{Optical spectra} 

\noindent{\bf SDSS:} 
The SDSS spectrum of J0746 is displayed in the top panel of
Figure~2. The data have a spectral resolution R $\sim$ 1900. 

The optical spectrum of J0746 is that of a typical $z \approx 3$
quasar. Broad prominent emission lines, such as Ly$\alpha$ (4827 \AA),
C~IV 1549~\AA\ (6164~\AA), and Si~IV/O~IV 1400~\AA\ (5571 \AA),
typical of high-$z$ quasars, are apparent in the spectrum. The
continuum power law slope in the SDSS spectrum is $\alpha{\nu} \sim
0.54$, very near the canonical $\alpha{\nu}=0.5$ from the SDSS quasar
composite (Vanden Berk et al. 2001).

We calculated the observed EWs and fluxes of the detected lines. For
C~IV we find EW$_{obs}$=31 \AA\ and F$_{\lambda} \sim 9.8 \times
10^{-16}$ \flux \AA$^{-1}$, while for Si~IV/O~VI EW$_{obs}$=78 \AA\
and F$_{\lambda} \sim 2.5 \times 10^{-15}$ \flux \AA$^{-1}$,
respectively. The EWs are within the distribution for radio-loud and
radio-quiet quasars (Wills et al. 1993). There is no evidence for C~IV
absorption, either in the emission line or in the neighboring
continuum.

\noindent{\bf HET:} Blazars have notoriously variable levels of
optical continuum which may dilute the emission lines during high
states. We thus acquired a second optical spectrum of J0746 in 2005
Fall with the Hobby-Eberly Telescope (Ramsey et al. 1994). The HET
spectrum of J0746 was obtained on 10 October 2005 with the Low
Resolution Spectrograph (Hill et al. 1998).  The 1.5\arcsec\ slit, 300
line mm$^{-1}$ grating, and GG385 filter produced a spectrum from
4000-8000~\AA\ at a resolution R=460.  The spectrum, which is the sum
of two 900~s exposures, is shown in the lower panel of Figure~2.

The HET spectrum, taken 3.7 years (only 11 rest-frame months) after
the SDSS one, shows very similar features. The HET continuum is redder
and the lines slightly stronger, but these changes should be
considered marginal at best; the quality of the spectrophotometry of
the SDSS data is considerably higher than that of the HET observation,
and the change in the equivalent widths is not much more than
1$\sigma$. Thus, we can conclude that the level of the optical
continuum did not significantly change between the epochs of the SDSS
and HET observations.

\subsection{Radio} 

The radio observations are presented in Table~3. 

\noindent{\bf UMRAO:} 
The University of Michigan radio data were obtained using the UMRAO
26-meter prime focus paraboloid antenna which is equipped with
wide-band room temperature HEMPT (High Electron Mobility Pseudomorphic
Transistor) amplifiers and operates at central frequencies of 4.8,
8.0, and 14.5~GHz. Measurements at all three of these frequencies
utilize rotating, dual-horn polarimeter feed systems. An on-off
observing technique was used at 4.8~GHz, and an on-on technique
(switching the target source between the two feed horns closely spaced
on the sky) was employed at 8 and 14.5~GHz. A typical observation
consisted of a series of 8 to 16 individual measurements over a 25 to
45 minute period, depending on the observing frequency.

Observations of the program source were preceded and followed by
observations of nearby calibrator sources (3C 144, 3C 145, 3C 218, 3C
274) selected from a grid across the sky; these calibration
measurements were used to correct for temporal changes in the antenna
aperture efficiency.  Frequent drift scans are also made across
stronger sources throughout every run to verify the telescope pointing
correction curves. The flux scale for the UMRAO observations is set by
observations of Cassiopeia~A (e.g., see Baars et al. 1977).  A more
detailed explanation of the calibration and analysis techniques is
given in Aller et al. (1985).  Interpolation of the radio fluxes in
Table~3 yields a slope $\alpha_r=0.08$, typical of a core-dominated
source as shown by the \vlba\ images at 2.3 GHz and 8.5 GHz (Fey \&
Charlot 2003). Comparing our 8~GHz measurements with the long-term
8~GHz light curve in Fig. 8 of Lazio et al. (2001) shows that at the
time of the \swift\ observations in November 2005 the source was in a
high intensity state in the radio, similar to the maximum recorded in
late 1992. 

\noindent{\bf Mets\"ahovi:} 
Observations were also made with the 14-m diameter Mets\"ahovi radio
telescope at the frequency of 37 GHz. The fluxes are listed in
Table~3. 

The 37 GHz observations were made with the 13.7 m diameter Mets\"ahovi
radio telescope, which is a radome enclosed paraboloid antenna
situated in Finland (24 23' 38''E, +60 13' 05''). The measurements
were with a 1 GHz-band dual-beam receiver centered at 36.8 GHz. The
HEMPT front end operates at room temperature. The observations are
ON--ON observations, alternating the source and the sky in each feed
horn. A typical integration time to obtain one flux density data point
is 1200--1400 s.

The flux density scale is set by observations of DR 21. Sources 3C 84
and 3C 274 are used as secondary calibrators. A detailed description
of the data reduction and analysis is given in Ter\"asranta et
al. (1998). The error estimate in the flux density includes the
contribution from the measurement rms and the uncertainty of the
absolute calibration.

\section{X-ray results} 

\subsection{Timing analysis} 

Here we discuss the results of a timing analysis of the XRT and BAT
data.  Variability was assessed through the use of a $\chi^2$ test,
with related probability P$_{\chi^2}$ that the light curve is
constant; thus, small values of P$_{\chi^2}$ indicate significant
variability.

\noindent{\bf BAT:} 
The BAT light curve in the energy range 15--195~keV is shown in
Figure~3a. Bins are 15-day averages. The vertical ticks mark the dates
of the three September XRT observations. The BAT light curve is
formally consistent with a constant, with P$_{\chi^2}$=0.12.

\noindent{\bf XRT:}
The XRT light curves were extracted in soft (0.5--2~keV), hard
(2--8~keV), and total (0.5--8~keV) energy bands. We searched for
spectral changes by computing hardness ratios defined as the ratio of
the hard and soft X-ray light curves. 

Figure~3b shows the XRT long-term light curve from 0.5--8~keV,
obtained from the average count rates at the four observing epochs in
Table~1. From Figure~3b it is apparent that the flux changed by a
factor \gtsima 2 on timescales of a week. The $\chi^2$ test confirms
variability, with P$_{\chi^2} < 10^{-5}$. No spectral changes were
observed, with the plot of the hardness ratios versus time being
constant (P$_{\chi^2} \sim 0.098$). 

Next, we searched for variability within the XRT individual
exposures. Figure~4 shows the 0.5--8~keV light curves at the four
epochs, rebinned at 2000~sec resolution. Flux variations are observed
at every epoch, and most notably during the Sept 5 and Nov 4
observations. A $\chi^2$ test for these two light curves yields
significant variability, with P$_{\chi^2} < 10^{-3}$ and P$_{\chi^2}
\sim 9 \times 10^{-5}$, respectively. At the beginning of the Sept 5
observation the flux increases by a factor 2.5 in one hour, albeit
within large uncertainties. A similar increase of the flux on a
timescale of $\sim$ 2 hrs is also present at the beginning of the Nov
4 observation. On this date, more modest variations are also seen
throughout the exposure, on similar timescales (1--2 ks).  No
significant spectral variability is detected in either the Sept 5 or
Nov 4 data, as shown by the the hardness ratio light curve
(P$_{\chi^2} > 0.06$).

We also investigated the energy-dependence of the variability by
calculating the variability amplitude relative to the mean count rate,
F$_{var}$, corrected for effects of random errors (e.g., Edelson et
al. 2002), for both hard and soft Nov 4 light curves. In practice,
however, due the paucity of data points and the large statistical
errors, only in few cases were we able to estimate F$_{var}$.  There
is a suggestion of a larger variability amplitude at harder X-rays,
with F$_{var, hard}$=0.160 and F$_{var, soft}$=0.05 for the hard,
although higher quality data are necessary to reach firmer conclusions.

\noindent{\bf Comparison to previous X-ray observations:} As mentioned
above, J0746 was detected during the RASS with a 0.2--2~keV count rate
of 0.04 c/s. To compare the \rosat\ and \swift\ observations, we
converted the RASS 0.2--2~keV count rate into an XRT 0.5--8~keV count
rate. Using \verb+PIMMS+ and assuming an absorbed power law with
column density N$_H=3 \times 10^{21}$ \nh\ and photon index
$\Gamma=1.3$ (see below), we find that the 0.5--8 keV count rate at
the time of the RASS observation was 0.041 c/s, comparable to the
lower state of the XRT observation in September 3 (Table~1). Thus, at
the time of the \rosat\ sky survey J0746 was in a lower X-ray
intensity state.

We conclude that the medium-hard X-ray emission of J0746 is
variable. There is a hint for larger-amplitude variations at the
harder energies, which will be tested in the \suzaku\ observation.

\subsection{Spectral analysis} 

\subsubsection{The X-ray continuum} 

The results of the spectral fits to the XRT and BAT spectra are
reported in Table~4. The best-fit parameters and their 90\%
uncertainties for one parameter of interest ($\Delta\chi^2$=2.7) are
listed, together with the observed fluxes and intrinsic
(absorption-corrected) luminosities. 

\noindent{\bf BAT:}
We fitted the BAT spectra with a single power law model without an
absorption column density, as the hard X-rays are not affected by the
relatively small column densities measured with the XRT (see
below). The BAT continuum from the first 3.5 months of the survey,
when the source was discovered, is fitted by a very hard power law,
$\Gamma \sim 0.74$. The isotropic X-ray luminosity of L$_{15-150~keV}
\sim 9 \times 10^{47}$ \lum\ makes J0746 one of the most luminous 
extragalactic sources at hard X-rays to-date.

The BAT continuum from the whole 9 months integration, however, is
described by a steeper power law, $\Gamma=1.3$. The derived flux is a
factor 2 lower than from the 3-months spectrum, implying that the
source was already fading after the first 3 months.  However, as shown
in the top panel of Figure~5, this is only a \gtsima 1$\sigma$
effect. The high-energy spectral variability properties of J0746 will
be better quantified in the \suzaku\ HXD observations. 


\noindent{\bf XRT:}
The XRT spectra were initially fitted with a single power law with the
Galactic column density. At all epochs, the residuals of this model
showed a deficit of flux below 3 keV, indicating the presence of
spectral flattening at these energies. This is illustrated in the
bottom panel of Figure~5, where the residuals of a power law with
Galactic N$_H$ from the joint fits to the 4 epochs is plotted. The
flattening below 3 keV can be interpreted as due either to excess
absorption over Galactic, or as a flatter power law component. We
tested both possibilities by fitting the XRT data with two models: 1)
a single power law plus an additional free column density, N$_H^z$, at
the source rest-frame; and 2) a broken power law with Galactic
absorption only.

For model 1, the choice of an intrinsic absorber is motivated by the
fact that no evidence for absorption due to an intervening medium is
present in the optical spectra of the quasar (Fig. 2). Moreover,
excess X-ray absorption along the line of sight was previously found
to be common in $z>1$ radio-loud quasars (see Worsley et al. 2006 and
references therein), but not in coeval radio-quiet sources, suggesting
the absorbing medium is located near the central engine of the AGN
(Elvis et al. 1992, Fiore et al. 1998, Page et al. 2005).

The fit with model 1 is reported in Table~4. In the three longest
observations (obsids 002, 003, 007), the addition of the intrinsic
absorber is significant at 99\% confidence ($\Delta\chi^2$=5) for 002
and 003, and at 92\% ($\Delta\chi^2$=4) confidence for 007, according
to the F-test.  The measured intrinsic column density is N$_H^z=(1-4)
\times 10^{22}$ \nh, and the photon index is $\Gamma=1.3$, at the flat
end of the distribution for $z>1$ radio-loud quasars with $\log R_L
\approx 3$ (Page et al. 2005). In the case of obsid 001, corresponding
to the lowest intensity state and shortest exposure time, the column
density and slope from a free-fit are poorly constrained; we thus
repeated the fit by fixing N$_H^z$ to the average value from the fits
to the remaining three observations. This yields $\Gamma \sim 1.3$,
similar to the other three exposures and confirming that the
0.5--8~keV flux changes with no accompanying spectral variations
(within 20\%).

Extrapolating the best-fit model of the November XRT observation to
energies $>$ 10~keV yields a F$_{15-150~keV}=2.4 \times 10^{-11}$
\flux. This is a factor 4 smaller than the flux from the 9-months
survey, when the BAT flux was already declining. It is thus likely 
that during the November 2005 campaign the source was in a low state.

Acceptable fits were obtained from model 2, a broken power law with
Galactic absorption; the parameters are reported in Table~4. Indeed,
in the case of the longest exposure of November, the broken power law
model is formally preferred (at 98\% confidence, $\Delta\chi^2$=8) to
model 1.
This is a well-known observed phenomenon in high-z radio-loud quasars:
the spectral flattening at soft energies can be equally well described
by a power law with absorption (either cold or warm) and a convex
broken power law. We will revisit this issue in \S~5.3. 

We thus conclude that the XRT spectra are well-described by a single
power law with strong spectral flattening below a few keV, which can
be described in terms of excess absorption over the Galactic value
(Table~4). Acceptable fits are also obtained with a convex broken
power law, yielding $\Delta\Gamma \sim 1.1$ at 4~keV. 

\subsubsection{Limits on the \feka\ emission line}

Usually, in blazars the beamed jet emission dilutes or overwhelms the
reflection features from the disk, yielding a featureless X-ray
spectrum (Sambruna, Chou, \& Urry 2000). However, during a low state
of the non-thermal emission the \feka\ line can be detected. For
example, in 3C~273, Grandi \& Palumbo (2004) observed a variable EW of
the \feka\ line and estimated that the X-ray disk spectrum is diluted
by the beamed radiation of the jet by a factor 1.2 to 2.8 from 
2--10~keV, while the jet always dominates above 40~keV. Similarly, a
\feka\ line with EW $\sim$ 100 eV was detected from the
cores of two powerful quasars hosting one-sided \chandra\ jets
(Sambruna et al. 2006, in prep.). 

We thus investigated the presence of an \feka\ emission line in J0746,
which in this source is redshifted to 1.61 keV, at the peak of the
detector effective area. We concentrated on the Nov 4 spectrum, which
has the best signal-to-noise ratio. An unresolved (width
$\sigma_l=0.05$ keV) Gaussian line with rest-frame energy fixed at
6.4~keV was added to the power law best-fit model.  The 90\%
confidence upper limit to the rest-frame EW is 200 eV.
 
No evidence for such a line is present in the other datasets. Adding a
Gaussian line at 1.61~keV to the power law model that best-fits the
four datasets jointly yields $\Delta\chi^2=-1$, indicating that the
Gaussian component is not needed.

\section{Discussion} 

\subsection{What is J0746?} 

Below we discuss the evidence supporting the classification of J0746
as a high luminosity blazar, or more precisely a Flat Spectrum Radio
Quasar (FSRQ), due to its prominent broad emission lines in the
optical. The high energy emission however can only be attributed to a
relativistic jet. 

\subsubsection{Multiwavelength Variability} 

One of the defining properties of blazars is their variability, often
dramatic, at all observed wavelengths from radio to gamma-rays (Ulrich
et al. 1995). Indeed, in J0746 flux and spectral variability is
observed in the radio and X-rays, the two bands with the best coverage
so far.

J0746 was observed as part of the Green Bank Interferometry monitoring
program at 2.3 and 8~GHz during the years 1979-1996 (Lazio et
al. 2001). Correlated flux and spectral variability was observed at
8~GHz, with a flux increase by a factor 3 in 3 years and steeper slope
with increasing flux. This is similar to other FSRQs in the Lazio et
al. sample. 

In the optical-UV, the \swift\ UVOT measurements are consistent with
the magnitudes from the SDSS, and with historical records
(\S~3.3). While this needs to be confirmed by future monitoring, the
lack of variability in this band is consistent with the idea that the
optical/UV radiation comes from a different component, most likely the
thermal emission from the disk. A straightforward test could be
provided by polarimetry, with a highly polarized optical flux arguing
against a thermal origin and in favor of a non-thermal one.  

J0746 was detected by the BAT at $>$ 10~keV with a bright flux. This
is similar to other blazars. Bright ($\sim 10^{-10}$ \flux) emission
was detected with the BAT (Giommi et al. 2006, submitted) and
\integral\ (Pian et al. 2006) from the $z$=0.859 FSRQ 3C~454.3, while
a rapid (lasting 2~ks) flare was detected with \integral\ in
20--40~keV at the position of the $z$=0.902 blazar NRAO~530 (Foschini
et al. 2006). Thus, it is reasonable to expect significant
shorter-term flux variability at energies $>$ 10~keV from J0746 in
future more sensitive observations. 

The X-ray variability is better constrained in the 0.5--8~keV energy
band. The XRT light curves of J0746 show flux changes of a factor 2 on
timescales of hours to weeks and months. There is a hint that the hard
X-rays vary with larger amplitude than the softer energies, but this
finding needs to be confirmed by future observations. However, while
the 0.5--8~keV flux varies, there are no accompanying spectral
variations. This is similar to other FSRQs at lower $z$ observed with
\asca\ and \rosat\ (Donato et al. 2001; Sambruna 1997), and to other
radio-loud quasars at $z>1$ (Page et al. 2005). 

The large X-ray luminosity and rapid variability of J0746 provide
further support for the idea that its emission is beamed. In fact,
assuming the X-ray emission from J0746 is isotropic, we can derive a
limit to the radiative efficiency $\eta$. Following Fabian (1979),
$\eta > 4.8 \times 10^{-43} \Delta L/\Delta t$ erg s$^{-1}$ s$^{-1}$,
where $\Delta L$ is the luminosity change in a time interval $\Delta
t$ (Brandt et al. 1999). From Table~4 and Figure 4d, $\Delta L \sim 5
\times 10^{46}$ \lum\ and $\Delta t \sim 1800$ sec in the quasar's
rest-frame. Thus, $\eta > 13$, which is unphysically large unless we
require relativistic beaming of the radiation.

\subsubsection{Broad-band spectral indices} 

The broad-band energy distributions of blazars can be described to 
first approximation with the spectral indices $\alpha_{ro}$,
$\alpha_{ox}$, and $\alpha_{rx}$. These are defined as the
two-point indices between radio (5~GHz) and optical ($V$ band),
optical and X-rays (1~keV), and radio to X-rays. Based on the
radio-to-X-ray flux ratio, blazars can be further classified as
High-Energy Peaked BL Lacs (HBLs), with $\alpha_{rx} < 0.8$; and
Low-energy Peaked BL Lacs (LBLs) and FSRQs, with $\alpha_{rx} > 0.8$
(Padovani \& Giommi 1995). 

When plotted in the $\alpha_{ro}- \alpha_{ox}$ plane, blazar classes
occupy distinct, although partly overlapping, regions. For HBLs,
$\alpha_{ro} < 0.6$ and $\alpha_{ox} < 1.3$, while LBLs and FSRQs have
$\alpha_{ro} > 0.5$ and $\alpha_{ox} > 1.0$ (e.g., Donato et
al. 2001). Using rest-frame flux densities at 5~GHz, in $V$ band, and
at 1~keV, the broad-band spectral indices of J0746 are
$\alpha_{rx}=0.78, \alpha_{ro}=0.65$, and $\alpha_{ox}=1.03$. Thus,
J0746 appears to be a borderline source between FSRQs and HBLs.

It is useful to compare the broad-band indices of J0746 to radio-quiet
quasars, because this could indicate the level of boosting of the
continuum due to beaming. A recent extensive compilation of broad-band
indices for radio-quiet quasars is provided by Strateva et
al. (2005). Here, they define an optical-to-X-ray index as the slope
between the flux at 2~keV and at 2500~\AA\ in the quasar's
rest-frame. For J0746, this index is 1.16, much flatter than the
radio-quiet quasars in Strateva et al. (2005). Thus, for a given
optical flux, the X-ray emission from J0746 is more than a factor 10
larger than expected for radio-quiet quasars. This also suggests that,
at least to a first approximation, beaming affects the X-ray emission
of J0746 but not (or much less) its optical-UV flux.

\subsubsection{The Spectral Energy Distribution (SED)} 

As discussed in \S~1, a strong spectral signature of blazars is the
presence of a double-peaked structure in their SEDs. Similarly, as
shown in Figure~6, J0746 exhibits a double-humped SED. 

To assemble the SED of J0746, we used data from this paper as well as
from the literature. In Figure~6 the radio (filled circles), optical,
UV, and X-ray (XRT, 0.5--8~keV) fluxes are contemporaneous
measurements from our November 4 campaign, while the filled triangles
are archival observations from \verb+NED+. Also plotted is the
9-months BAT spectrum (see below). As such, the SED in Figure~6
contains contemporaneous (radio, optical, UV, and medium-hard X-rays)
data as well as archival fluxes.

At GeV energies, we plot an upper limit to the EGRET flux from a
reanalysis of four archival images. Unfortunately, the coverage of the
J0746 field by EGRET was very poor: EGRET observed the source in 1992,
1993, 1994, and 1995. However, J0746 was never closer than 16.7\deg\
to the EGRET axis; this implies that no high-quality exposure of the
source was ever acquired. The upper limit to the $>$ 100 MeV flux from
the combined four observations is $5\times 10^{-8}$ ph cm$^{-2}$
s$^{-1}$ and is plotted in Figure~6 with an arrow.

The first striking feature in Figure~6 is that, as implied by the hard
BAT spectrum and the EGRET limit, the peak of the second spectral
component is located around a few MeV. This would make J0746 a new
member of the poorly known MeV blazars subclass, which so far contains
only two sources detected by COMPTEL (Blom et al. 1996; for PKS
0208--512 see, however, Stacy et al. 2003). The obvious caveat is, of
course, that the BAT and EGRET fluxes are not simultaneous, while
FSRQs notoriously vary at high energies on timescales of days.

The second important feature in Figure~6 is the steep optical-to-UV
continuum, which implies a peak of the synchrotron component at IR
wavelengths. Observations of J0746 in the near and far-IR are urged to
better sample the SED in this critical portion of the spectrum. It is
tempting to interpret the steep optical-UV emission as the high-energy
tail of thermal emission from the accretion disk, i.e., the so-called
blue bump, usually observed in quasars (Elvis et al. 1994) but also in
a few FSRQs (3C~273; von Montigny et al. 1997).

In conclusion, based on multiwavelength evidence, we classify J0746
with a high-$z$ blazar. Further observations, especially polarimetry
at radio through optical wavelengths, are strongly encouraged to
confirm the blazar nature of this source. 

Below we model the SED of J0746 in the assumption that J0746 is a
blazar, and that the optical-to-UV continuum at the time of the UVOT
observations is due to thermal emission from the disk.

\subsection{Modeling the Spectral Energy Distribution} 

The continuum from blazars is generally attributed to synchrotron and
inverse-Compton (IC) emission (e.g., Ghisellini et al. 1998; for a
different view see Atoyan \& Dermer 2003 and references therein)
produced within a jet pointing toward the observer. In the case of
blazars showing strong emission lines, as in J0746, the target photons
for the IC process are likely dominated by those belonging to the
external radiation field (disk, broad line region), amplified by
relativistic effects in the plasma rest-frame (Dermer \& Schlickeiser
1993, Sikora et al. 1994). Synchrotron self-Compton emission (Maraschi
et al. 1992), on the other hand, can substantally contribute in the
soft X-ray band (e.g., Ballo et al. 2002).

Again, we caution that the SED in Figure~6 contains non-simultaneous
data. In particular, the BAT and XRT spectra were taken several months
apart, while the BAT spectrum was integrated over a long-time period
(9 months). The choice to use the 9-months BAT spectrum in the SED
modeling is due to the fact that it is ``closer'' in time to the Nov 4
campaign data; indeed, the extremely flat continuum during the initial
higher state, $\alpha \sim -0.26$, posits severe problems for current
blazar models.

To reproduce the observed SED we have applied the synchrotron-IC
emission model fully described in Maraschi \& Tavecchio
(2003). Briefly, the emission region is modeled as a sphere with
radius $R$, in motion with bulk Lorentz factor $\Gamma$ at an angle
$\theta $ with respect to the line of sight, and filled with tangled
magnetic field (with intensity $B$) and relativistic
electrons. $\Gamma$ and $\theta$ are combined in the Doppler factor
$\delta=[\Gamma (1-\beta \cos \theta)]^{-1}$, where $\beta=v/c$.  The
(purely phenomenological) electron distribution is modeled as a
smoothed broken power law, with indices $n_1$ and $n_2$ below and
above the break Lorentz factor $\gamma_b$. The electron distribution
extends within the limits $\gamma_{\rm min} <\gamma <\gamma_{\rm
max}$.

The exceptionally steep optical-UV continuum strongly suggests that
the emission in this band is dominated by a thermal-like component,
possibly associated with the blue-bump component which characterizes
the quasar optical region (e.g., Sun \& Malkan 1989), usually thought
to be associated with the putative accretion disk. This interpretation
is justified by the presence of prominent emission lines in the HET
spectrum of J0746 (\S~3.2), which was taken a few weeks after the end
of the BAT observations discussed here. 

The identification of the optical-UV component with the disk emission
constrains the disk luminosity L$_D \sim$ a few $\times 10^{47}$
\lum. We choose to model this component with a black body peaking at
the typical frequency of $10^{15}$ Hz (in the quasar rest-frame). The
external radiation field is modelled as the same black-body, diluted
within the broad line region, supposed to be spherical with radius
$R_{\rm BLR}$. This radiation approximates the radiation reprocessed
and re-isotropized by the BLR clouds. Its luminosity is fixed to a
fraction $\tau $ of the luminosity $L_D$ of the putative central
illuminating disk.

The resulting model is reported in Figure~6 (solid line), with the
parameters listed in the caption. From the inferred parameters and
assuming the composition of one proton per emitting electron (for the
reason of this choice see Maraschi \& Tavecchio 2003), we can infer a
kinetic power of the jet of $P_j=4\times 10^{48}$ \lum. This can be
compared with the radiative output of the disk $L_{\rm D} = 4\times
10^{47}$ \lum. The ratio $P_j/L_D\sim 10$ is of the same order of what
is found in other cases for which both powers can be estimated with
reasonable accuracy (Tavecchio et al. 2000; Maraschi \& Tavecchio
2003). The parameters used to reproduce the SED are close to those
usually found for other powerful blazars (e.g., Maraschi \& Tavecchio
2003, Ghisellini et al. 1998).

An interesting aspect of the \swift\ results is the exceptionally flat
spectrum measured in the hard X-rays. Flat spectra ($\alpha_X = 0.3$)
characterize several high-luminosity blazars (e.g., Tavecchio et
al. 2000). To model such hard spectra we use an electron index of
$n_1=1.5$. Such a flat electron distribution poses serious problems
for the standard scenarios usually discussed, involving shock
acceleration or cooling processes, which lead to a typical
distribution with $n=2$. An interesting possibility has been
considered by Sikora et al. (2002). They assume a two-step
acceleration process: the flat portion is produced by a
pre-acceleration mechanism (e.g., stochastic acceleration driven by
turbulence or reconnection), while the steep portion after the peak is
produced through the standard shock acceleration.

In the radio band, our data fits well to a power law spectrum with
$\alpha=0.48$, quite typical of a transparent synchrotron source.
Without a high resolution VLBI image, the most straight forward
interpretation is that there is an extended region around the central
core that is responsible for the radio-mm flux.  The IR-optical
emission is presumably produced in a core region that starts appearing
at shorter wavelengths, but without data in the sub-mm it is not
possible to specify exactly where the BLR region starts to dominate.
The discontinuity between the radio-mm spectrum and the shorter
wavelength emission is consistent with a high energy cutoff in the
radiating particles in this extended emitting region; such a cutoff is
already part of the model parameters used to fit the core region at
shorter wavelengths.

It is worth noting that we fix $\gamma_{\rm min}$ to 1. Larger
values would in fact produce an unobserved deficit of soft X-ray
photons, further flattening the X-ray spectrum. This is also typical
of this kind of sources, and seems to support the two-step view, in
which the bulk of the electrons remains at low energies.

Finally, while we have assumed that the steep optical-UV continuum in
Figure~6 is thermal in origin and related to the accretion disk, we
can not exclude {\it a priori} that the jet does not contribute to a
fraction of the optical flux, or even dominates it during
outbursts. In this sense, variability of the optical-UV continuum in
J0746 would have to be modeled as a mix of a non-thermal, variable jet
component and a thermal, less variable one related to the accretion.

\subsection{J0746 in context} 

There is evidence that the double-humped SEDs of blazars form a
continuous sequence in luminosity (Sambruna, Maraschi, \& Urry
1996). In this sequence, going from the more luminous FSRQs to the
fainter and closer TeV BL Lacs, the synchrotron component peak shifts
from IR to X-ray wavelengths, while the Compton peak shifts, by the
same amount, from GeV to TeV gamma-rays (Fossati et al. 1998). Models
of the SED sequence suggest that particle energies decrease and
magnetic fields increase with increasing luminosities while the
Doppler factor is approximately constant (Ghisellini et al. 1998). The
sequence finds a physical basis in the run of the jet and disk powers,
P$_{jet}$ and P$_{disk}$, with P$_{jet} \sim$ P$_{disk} \sim
10^{46}-10^{47}$ \lum\ for FSRQs and P$_{jet} >$ P$_{disk}$ for BL
Lacs (Tavecchio et al. 2000; Maraschi \& Tavecchio 2003). 


To compare the SED of J0746 to other blazars, we have plotted in
Figure~7 the blazar luminosity sequence from the reanalysis of Donato
et al. (2001). The solid line represents the best-fit model to the SED
of J0746 from Figure~6. It is apparent that the spectral properties of
this source are extreme in the blazar class: as discussed above, the
synchrotron peak likely lies below the IR frequency range while the IC
peak is at MeV energies. This qualifies J0746 as a member of the
poorly known ``MeV blazars'' subclass (Blom et al. 1996).

There is much debate whether the blazar sequence is intrinsically true
or a product of several observational biases, especially at gamma-rays
(Urry 1999), or limited sensitivity of current radio and optical
surveys (Nieppola, Tornikoski, \& Valtaoja 2006; Anton et
al. 2004). Nevertheless, here we used the sequence as a comparison
template to showcase the unusual properties of J0746. Note that, if
the sequence is true, the SED of J0746 would be extreme, but not
unexpected. 

A property J0746 shares with other high-$z$ radio-loud quasars is the
presence of spectral hardening in the X-ray spectrum below a few
keV. As discussed in \S~4.2.1, this can be modeled in terms of either
excess absorption along the line of sight, or a flatter power law. We
now discuss the two scenarios in turn. 

In the context of blazar emission models, a continuum break at soft
X-rays can be expected if the particle distributions responsible for
the high-energy component have a low-energy cutoff, parameterized by
the value of the minimum Lorentz factor of the relativistic electrons
(Tavecchio et al. 2000). In this scenario, the hardening occurs below
$\sim \gamma_{min}^2 \Gamma^2 \nu_{ext}$, with $\nu_{ext} \sim
10^{15}$ Hz (Fig. 6). The fitted value of the break energy in J0746
would thus imply $\gamma_{min} \sim 3$, consistent with the SED
modeling (Fig. 6). 


Alternatively, the spectral flattening below 3 keV can be described in
terms of excess N$_H$ along the line of sight, as originally advocated
for high-z radio-loud quasars by Elvis et al. (1992). Recent \xmm\
observations have confirmed that at least a fraction of these sources
could have intrinsic absorption, with N$_H^z \sim 10^{22}-10^{23}$
\nh\ (see Worsley et al. 2006 and references therein). The nature and
location of the absorber, however, is still debated today. The large
X-ray column densities would imply large optical reddening for the
quasars, which is not seen.

Such difficulty exists for J0746 as well. The SDSS and HET spectra
indeed show little or no intrinsic reddening, at odds with the X-ray
column density N$_H \sim 10^{22}$ \nh\ which implies $A_V \sim 10$ 
(assuming Galactic dust-to-gas ratios). Moreover, while several X-ray
absorbed quasars, both radio-loud and radio-quiet, show accompanying
UV absorption lines (e.g., Brandt, Laor, \& Wills 2000), this is not
the case for J0746 (Fig. 2). This would seem to indicate a dust-free,
neutral absorber. However, perhaps the major difficulty is to devise a
physically plausible location for the absorber. As shown in Figure~6
the X-ray emission from J0746 comes from a highly beamed ($\delta \sim
20$) jet, at least in its inner regions. One can thus speculate that
the absorbing medium is located inside the jet, perhaps ambient gas
entangled in the moving blobs (e.g., Celotti et al. 1998). 
However, whether this gas has a covering factor large enough to
produce significant spectral depression in the X-ray continuum
(Fig.~5) remains to be proven.

In conclusion, the nature of the low-energy spectral flattening in
J0746 is still an open question. We look forward to the deeper 
(100~ks), broad-band \suzaku\ spectrum which will hopefully settle the
issue.

\section{Summary and Conclusions} 

The BAT experiment onboard \swift\ detected bright emission in the
energy range 15--150~keV from the radio-loud quasar J0746 at
$z$=2.979, implying an isotropic luminosity around $\sim 10^{48}$
\lum. Rapid variability in 0.5--10~keV strongly suggests beaming. 
In this paper, we reported the \swift\ discovery observations and
follow-up multiwavelength monitoring of this source. The main results
are:

\begin{itemize} 

\item The X-ray emission of J0746 is variable on timescales 
of days to months (Fig. 3, 4). Fast variability on timescales of
1--2~ks is observed for the 0.5--8~keV flux (Fig.~3b). 

\noindent At all epochs, the XRT hardness ratio light curve is
consistent with a constant value, implying there is no spectral
variability on short (hours) timescales in 0.5--8~keV.

\item The 15--150~keV emission is consistent with a very hard 
continuum, $\Gamma \sim 0.74$, during the first 3 months of the BAT
survey, and a steeper continuum, $\Gamma \sim 1.3$, with halved flux, 
in the BAT spectrum integrated over 9-months. 

\item The 0.5--8~keV spectrum is well described by a $\Gamma \sim 1.3$
power law. Spectral flattening is present at $<$ 1 keV, which can be
interpreted as excess absorption over the Galactic value, implying an
intrinsic column density N$_H^z \sim 10^{22}$ \nh, or a flatter power
law component, implying a sharp ($\Delta \Gamma \sim 1.1$) spectral
break around 16~keV in the quasar's rest-frame.

\item The Spectral Energy Distribution of J0746 has a 
double-humped structure, with the first component peaking around IR
wavelengths and the second component likely at a few MeV. The
luminosity of the source is dominated by the high-energy emission. 

\end{itemize} 

Based on the results presented in this paper, we conclude that J0746
is a blazar, and more precisely, a Flat Spectrum Radio Quasar. Its SED
has extreme, but not unexpected, properties compared to other
blazars. Additional and possibly simultaneous multiwavelength
observations of J0746 are encouraged to confirm the jet-origin of its
emission; to this end, polarization measurements at radio and
optical wavelengths will be especially revealing.

The detection of a flare at high-energies from J0746 raises several
outstanding questions. How common are sources like J0746? What are the
physical parameters of MeV blazar jets? What causes the violent
outbursts at hard X-rays, and what is their duty cycle? The latter
issue is particularly important in view of the possibility that
blazars are significant, and even dominant, contributors to the
gamma-ray sky (Salamon \& Stecker 1994). The BAT flux variability on
timescales of months, as well as the BAT detection of several other
sources with properties similar to J0746 (Sambruna et al., in prep.),
suggests that flares at energies $>$ 10 keV may be common from these
blazars, stressing the key importance of wide-angle, sensitive and
continuous sky monitoring at these energies. 

\acknowledgements

This research has made use of data obtained from the High Energy
Astrophysics Science Archive Research Center (HEASARC), provided by
NASA's Goddard Space Flight Center, and of the NASA/IPAC Extragalactic
Database (NED) which is operated by the Jet Propulsion Laboratory,
California Institute of Technology, under contract with the National
Aeronautics and Space Administration. UMRAO is partially funded by a
series of grants from NSF and NASA and by the University of
Michigan. This work was supported in part by National Science
Foundation grant AST-0307582 (DPS) and NASA LTSA grant NAG5-13035
(WNB, DPS). 

Funding for the creation and distribution of the SDSS Archive has been
provided by the Alfred P. Sloan Foundation, the Participating
Institutions, the National Aeronautics and Space Administration, the
National Science Foundation, the U.S.  Department of Energy, the
Japanese Monbukagakusho, and the Max Planck Society.  The SDSS Web
site \hbox{is {\tt http://www.sdss.org/}.}  The SDSS is managed by the
Astrophysical Research Consortium (ARC) for the Participating
Institutions.

The Hobby-Eberly Telescope (HET) is a joint project of the University
of Texas at Austin, the Pennsylvania State University, Stanford
University, Ludwig-Maximillians-Universit\"at M\"unchen, and
Georg-August-Universit\"at G\"ottingen.  The HET is named in honor of
its principal benefactors, William P. Hobby and Robert E. Eberly.  The
Marcario Low-Resolution Spectrograph is named for Mike Marcario of
High Lonesome Optics, who fabricated several optics for the instrument
but died before its completion; it is a joint project of the
Hobby-Eberly Telescope partnership and the Instituto de
Astronom\'{\i}a de la Universidad Nacional Aut\'onoma de M\'exico.



\scriptsize
\begin{center}
\begin{tabular}{lllllll}
\multicolumn{7}{l}{{\bf Table 1: X-ray Observations}} \\
\multicolumn{7}{l}{   } \\ \hline
& & & & & & \\
Instrument & Obs ID & Date & Start Time & End Time & Exposure & Count rate \\
& & & & & & \\
(1) & (2) & (3) & (4) & (5) & (6) & (7) \\
& & & & & & \\ \hline
& & & & & & \\
BAT & $\cdots$ & survey & 12/04/04 & 09/30/05 & 755,000 & (6.2 $\pm$ 7.2)$\times 10^{-4}$\\
XRT & 00035200001 & Sept 3 & 07:17:36 & 12:14:59 &  1928 & 0.0456 $\pm$ 0.006 \\  
    & 00035200002 & Sept 5 & 04:13:33 & 09:15:58 & 2712 & 0.0667 $\pm$ 0.005 \\
    & 00035200003 & Sept 9 & 01:29:31 & 22:33:58 & 3210 & 0.0879 $\pm$ 0.005 \\ 
    & 00035200007 & Nov 4  & 00:39:56 & 23:29:58 & 15,356  & 0.0613 $\pm$ 0.002 \\
& & & \\ \hline

\end{tabular}
\end{center}

\noindent
{\bf Explanation of Columns:} 1=Detector name; 2=Observation ID;
3=Date of observation in 2005, except for the BAT where the data are
from the 9-months survey; 4=Start time (UT); 5=End time (UT); 6=Exposure 
time in sec; 7=Background-subtracted count rate in 15-150 keV for BAT 
and in 0.5--8~keV for the XRT. 

\normalsize


\clearpage

\scriptsize
\begin{center}
\begin{tabular}{llllll}
\multicolumn{6}{l}{{\bf Table 2: Optical Observations}} \\
\multicolumn{6}{l}{   } \\ \hline
& & & & & \\
Filter & Wavelength & Date & Mag & Flux & Exposure \\
& & & & & \\
(1) & (2) & (3) & (4) & (5) & (6) \\
& & & & & \\ \hline
\multicolumn{6}{l}{{\bf Swift UVOT}} \\
& & & & & \\
V & 5460 & MJD 53678 & 18.95 $\pm$ 0.19 & 92.7 $\pm$ 16.2 & 1382 \\
B & 4350 &           & 19.91 $\pm$ 0.17	& 48.2 $\pm$ 0.8 & 915 \\
U & 3450 & 	     & 19.88 $\pm$ 0.17 & 17.8 $\pm$ 0.3 & 1235 \\
W1& 2600 &           & $>$ 20.32        & $>$ 8.8 & 2572 \\
M2& 2200 &           & $>$ 21.49	& $>$ 3.2 & 4028 \\
W2& 1930 &           & $>$ 21.41	& $>$ 3.4 & 5584 \\ 
& & & & & \\
V & 5460 & MJD 53679 & 19.35 $\pm$ 0.16 & 64.1 $\pm$ 0.9 & 2123 \\
B & 4350 &           & 19.42 $\pm$ 0.12	& 75.6 $\pm$ 0.8 & 1217 \\
U & 3450 & 	     & 19.97 $\pm$ 0.17 & 16.4 $\pm$ 0.3 & 2123 \\
W1& 2600 &           & $>$ 21.63        & $>$ 2.6       & 4251 \\
M2& 2200 &           & $>$ 22.22	& $>$ 1.6       & 6978 \\
W2& 1930 &           & $>$ 22.64	& $>$ 1.1       & 8491 \\
& & & & & \\ \hline
\multicolumn{6}{l}{{\bf SDSS}} \\
& & & & & \\
u & 3571 & Dec 19, 2001 & 20.76 & 18.0 & \\
g & 4653 &              & 19.71 & 47.4 & \\  
r & 6147 &              & 19.50 & 57.5 & \\
i & 7461 &              & 19.42 & 61.9 & \\
z & 8904 &              & 19.48 & 58.6 & \\
& & & & & \\ \hline 

\end{tabular}
\end{center}

\noindent
{\bf Explanation of Columns:} 1=Filter; 2=Center wavelength in \AA; 3=Date of
observation; 4=Observed magnitude; 5=Derived flux in $\mu$Jy, 
dereddened for Galactic absorption; 6=Exposure in sec.

\normalsize

\clearpage

\scriptsize
\begin{center}
\begin{tabular}{llll}
\multicolumn{4}{l}{{\bf Table 3: Radio Observations}} \\
\multicolumn{4}{l}{   } \\ \hline
& & & \\
Date & Freq.& Flux & Unc. \\
& & & \\
(1) & (2) & (3) & (4) \\
& & & \\ \hline
\multicolumn{4}{c}{{\bf UMRAO }} \\ \hline 
& & & \\
09/25 &  8.0 &   0.52 &  0.02 \\
10/07 &  8.0 &   0.79 &  0.07 \\
10/08 &  8.0 &   0.65 &  0.01 \\
10/20 & 14.5 &   0.49 &  0.06 \\
11/04 &  8.0 &   0.68 &  0.03 \\
11/05 &  8.0 &   0.66 &  0.05 \\
11/08 &  4.8 &   0.82 &  0.02 \\
11/10 &  4.8 &   0.81 &  0.03 \\
11/11 &  4.8 &   0.86 &  0.02 \\
11/12 &  8.0 &   0.64 &  0.03 \\
& & & \\ \hline
\multicolumn{4}{c}{{\bf Mets\"ahovi}} \\ \hline 
& & & \\
11/02 & 37 & 0.36 & 0.09 \\
11/02 & 37 & 0.44 & 0.06 \\
11/02 & 37 & 0.49 & 0.08 \\
11/04 & 37 & 0.48 & 0.11 \\ 
11/04 & 37 & 0.58 & 0.08 \\ 
& & & \\\hline 

\end{tabular}
\end{center}

\noindent
{\bf Explanation of Columns:} 1=Date (MM/DD) of 2005; 2=Frequency in GHz; 
3=Flux density in Jy; 4=Uncertainty on the flux density. 

\normalsize

\clearpage

\scriptsize
\begin{center}
\begin{tabular}{llllll}
\multicolumn{6}{l}{{\bf Table 4: Spectral Fitting Results}} \\
\multicolumn{6}{l}{   } \\ \hline
& & & & & \\
Obs. & N$_H^z$ & $\Gamma$ & $\chi^2$/dofs & Flux & Lum \\
& & & & & \\
(1) & (2) & (3) & (4) & (5) & (6) \\
& & & & & \\ \hline
\multicolumn{6}{c}{{\bf BAT }} \\
& & & & & \\
3 months  & $\cdots$ & 0.74$^{+0.29}_{-0.72}$ & 0.33/2 & 142 & 135 \\
9 months  & $\cdots$ & 1.34$^{+0.40}_{-0.39}$ & 0.18/2 & 79.9 & 189 \\
& & & & & \\
\multicolumn{6}{c}{{\bf XRT$^a$}} \\
& & & & & \\ 
00035200001 & $40^{+81}_{-38}$ & 2.37$^{+1.63}_{-1.24}$ & 1.32/3 & 2.2 & 32 \\
            & 3.4 fix & 1.25$^{+0.41}_{-0.33}$ & 1.56/4 & 2.6 & 7.2 \\ 
00035200002 & $4.2^{+4.7}_{-3.2}$ & $1.30^{+0.27}_{-0.25}$ & 1.20/17 & 4.1 & 12.1 \\
00035200003 & $4.7^{+4.1}_{-2.9}$ & 1.38$^{+0.26}_{-0.22}$ & 0.72/11 & 4.8 & 16.7 \\ 
00035200007 & $1.4^{+1.5}_{-1.1}$ & 1.31 $\pm$ 0.11 & 1.3/43 & 3.1 & 9.7 \\ 
            & 4.04fix & $\Gamma_1^b=1.04^{+0.11}_{-0.17}$ & 1.15/42 & 2.8 & 8.0 \\
            &         & $\Gamma_2^c=2.14^{+1.17}_{-0.66}$ &   &   & \\
& & & & & \\ \hline

\end{tabular}
\end{center}

\noindent
{\bf Explanation of Columns:} 1=Observation ID (see Table1); 2=Column
density in $10^{22}$ \nh\ at the redhsift of the source; 3=Photon index;
4=Reduced $\chi^2$ and degrees of freedom (dofs); 5=Observed 
20--200 keV (BAT) and 
2--10~keV (XRT) flux in $10^{-12}$ \flux; 
6=Intrinsic (absorption-corrected) 20--200 keV (BAT) and 
2--10~keV (XRT) luminosity in $10^{46}$ \lum. 

{\bf Notes:} a=A column density fixed to the Galactic value ($4.04 \times 10^{20}$ 
\nh) was included in all the fits. b=Photon index below 
the break energy E$_b=4.0^{+0.9}_{-2.5}$~keV from a fit with a broken
power law model; c=Photon index above the break energy
E$_b=4.0^{+0.9}_{-2.5}$~keV from a fit with a broken power law model.

\normalsize

\newpage


\vspace{-1.0cm}
\begin{figure}[h]
\begin{center}
\hbox{
\psfig{figure=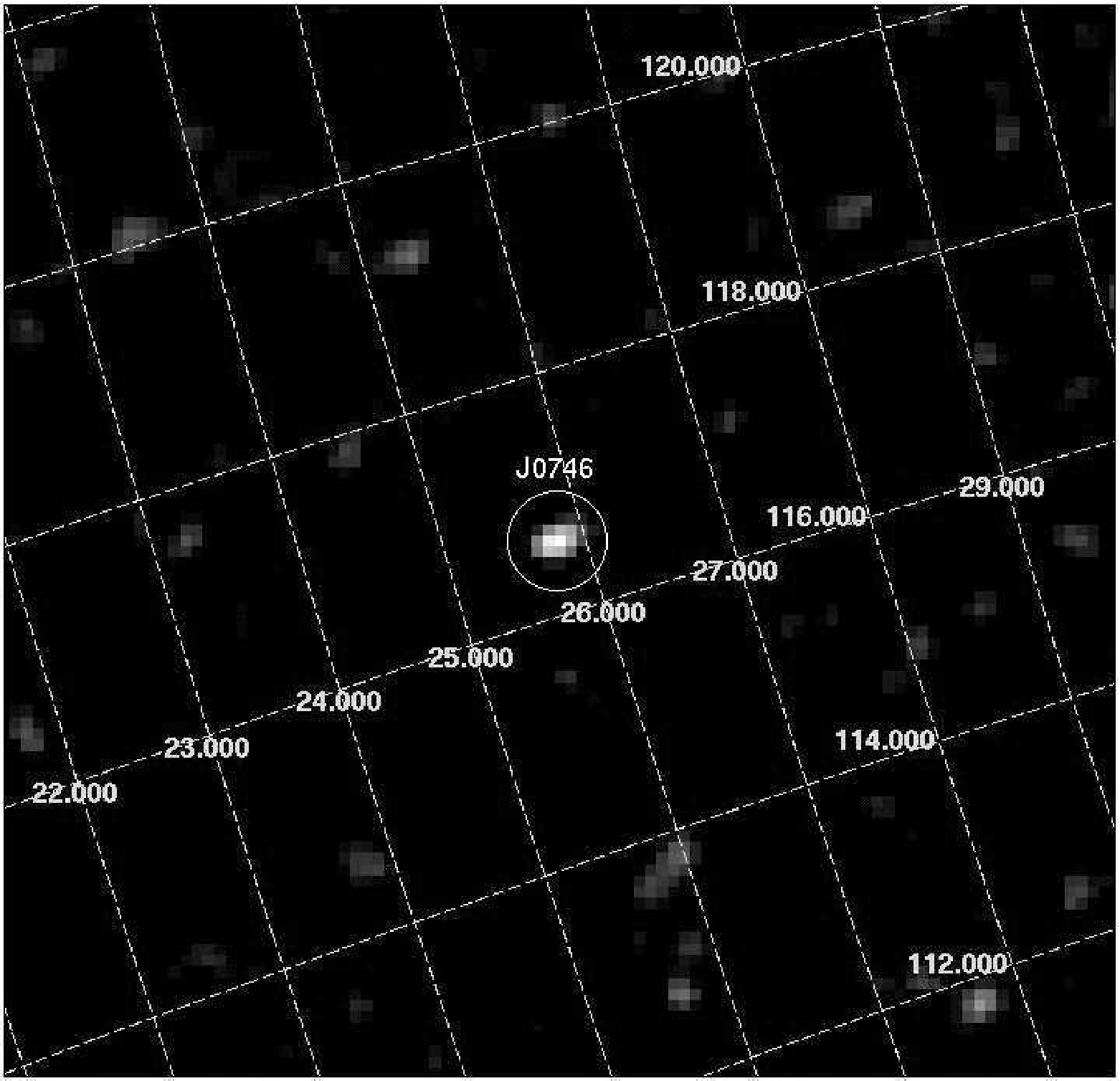,height=11cm,width=10cm}
}
\end{center}
\vspace{-1.0cm}
\caption{
{The BAT sensitivity image of J0746, ranging from $2\sigma$ (white)
to $7.5\sigma$ (black) for a total exposure time of 755~ks. 
The image is 10\deg x 10\deg, with 5\arcmin\
pixels.  The circle marks the position of J0746.  The orientation 
is arbitrary. J0746 is detected at $7.5\sigma$. 
}}

\end{figure}

\newpage


\vspace{-1.0cm}
\begin{figure}[h]
\begin{center}
\hbox{
\psfig{figure=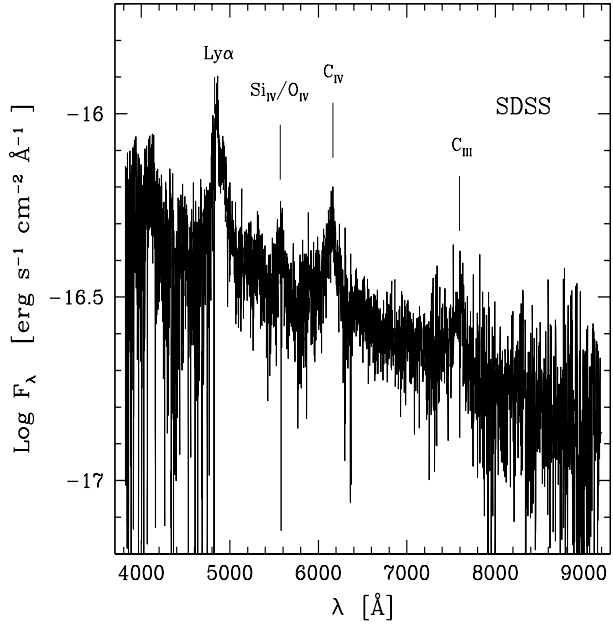,height=10cm,width=9cm}
}
\vspace{.2cm}
\hbox{
\psfig{figure=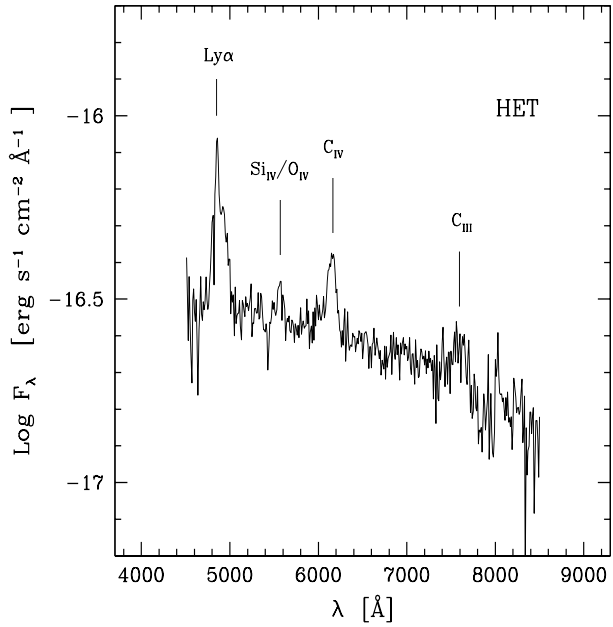,height=10cm,width=9cm}
}
\end{center}
\vspace{-1.7cm}
\caption{\scriptsize
{The SDSS {\bf (Top)} and HET {\bf (Bottom)} spectra of J0746, showing
typical broad UV lines of a high-redshift quasar. The SDSS spectrum
was taken December 19, 2001 and has a resolution R=1900; the HET
spectrum, taken on October 10, 2005, has a resolution R=460.}}

\end{figure}
\normalsize

\newpage


\begin{figure}[h]
\begin{center}
\hbox{
\psfig{figure=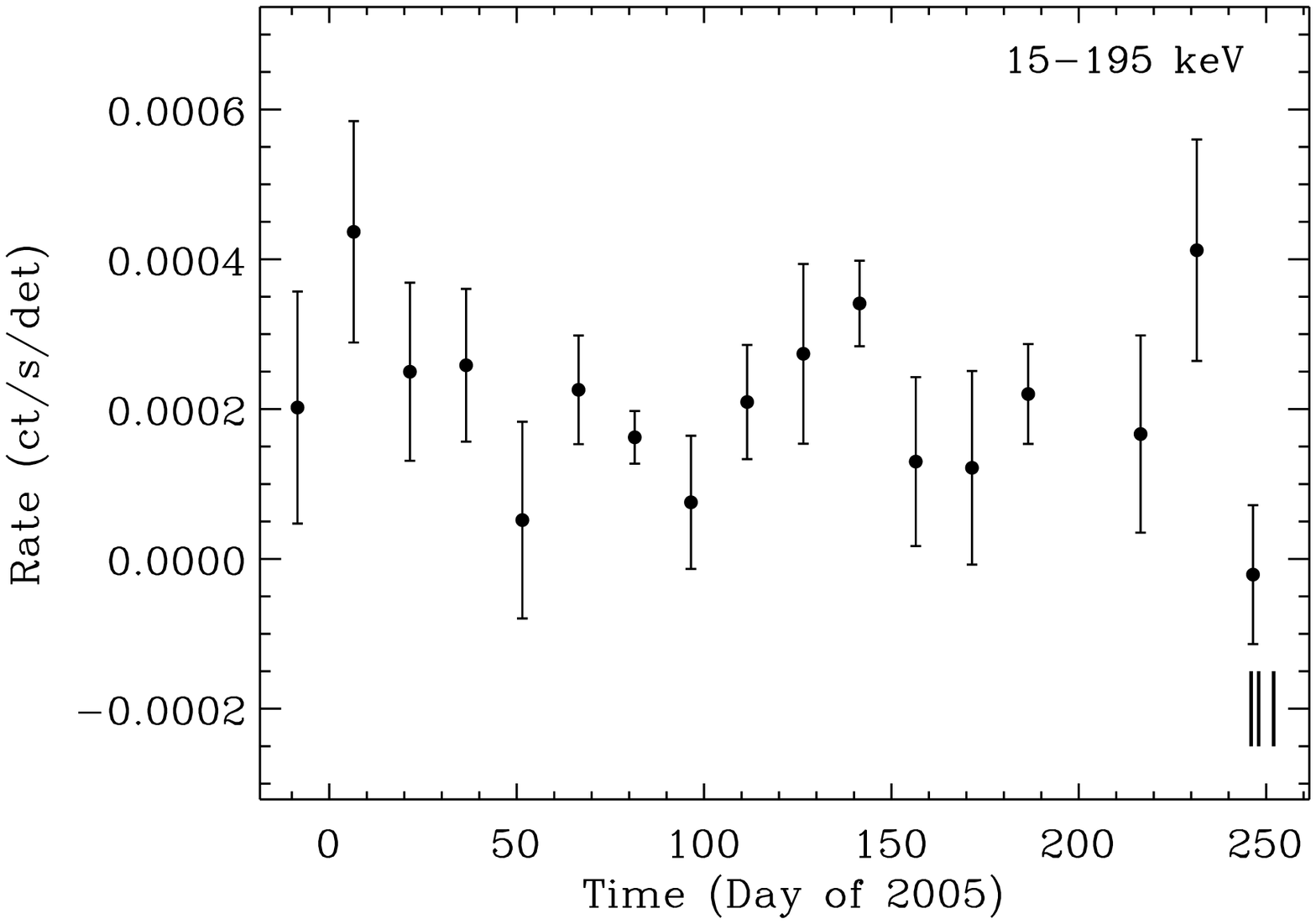,height=7.0cm,width=7.0cm}


\vspace{1.0cm}
\psfig{figure=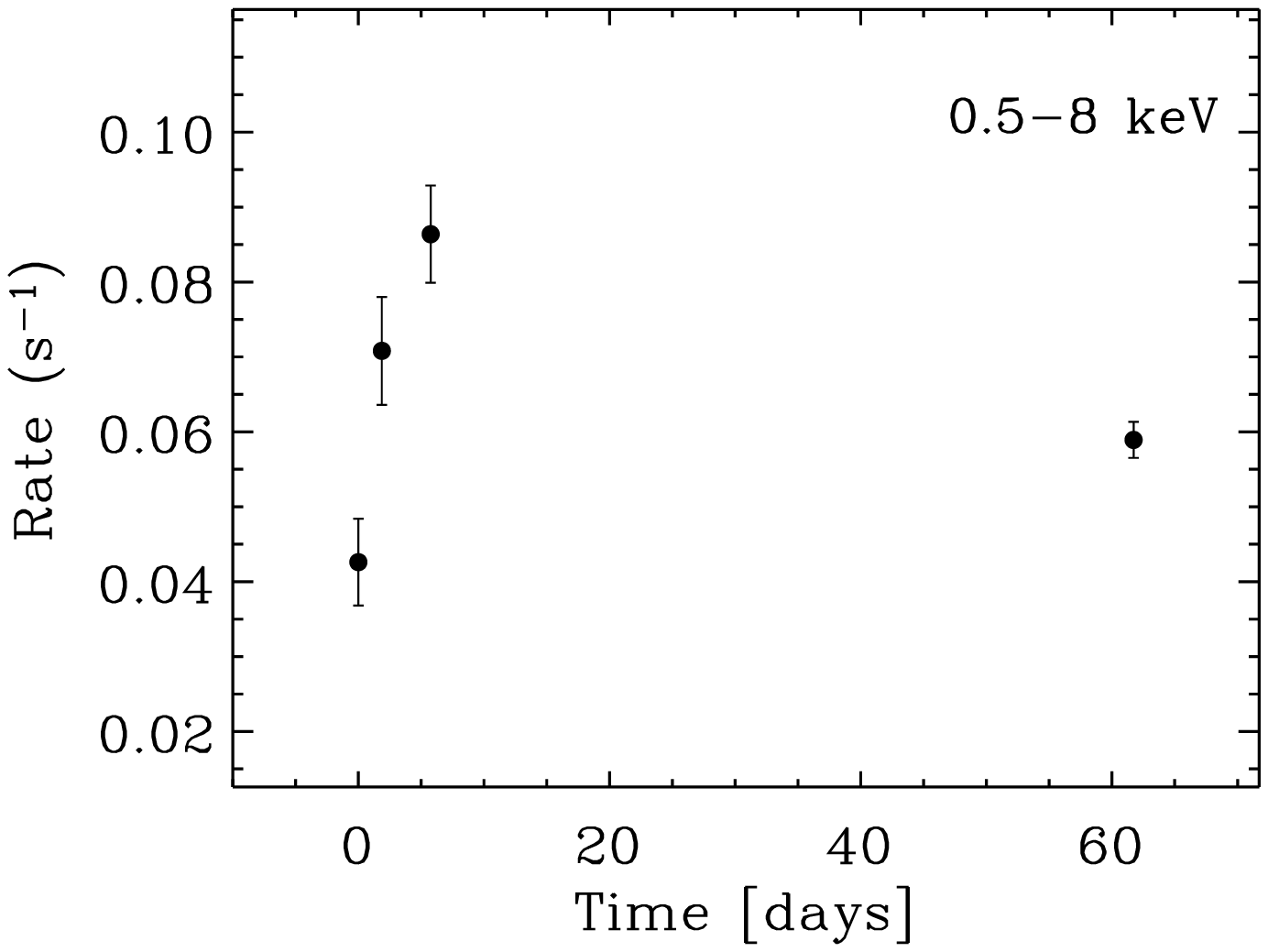,height=7.0cm,width=7.0cm}
}
\end{center}
\vspace{-1.0cm}
\caption{
{{\bf (a, Left):} The BAT light curve of J0746 from 15--195~keV. Bins
are 15-day averages. The start date is December 15, 2004. The
vertical bars in September mark the times of the follow-up XRT
observations (Table~1). {\bf
(b, Right):} The \swift\ XRT light curve, obtained by averaging the
count rate for each observing epoch (Table~1). Variation of the 0.5--8
keV flux of a factor \gtsima 2 in 5 days are apparent. }}

\end{figure}

\newpage


\begin{figure}[h]
\begin{center}

\hbox{


\vspace{1.0cm}
\psfig{figure=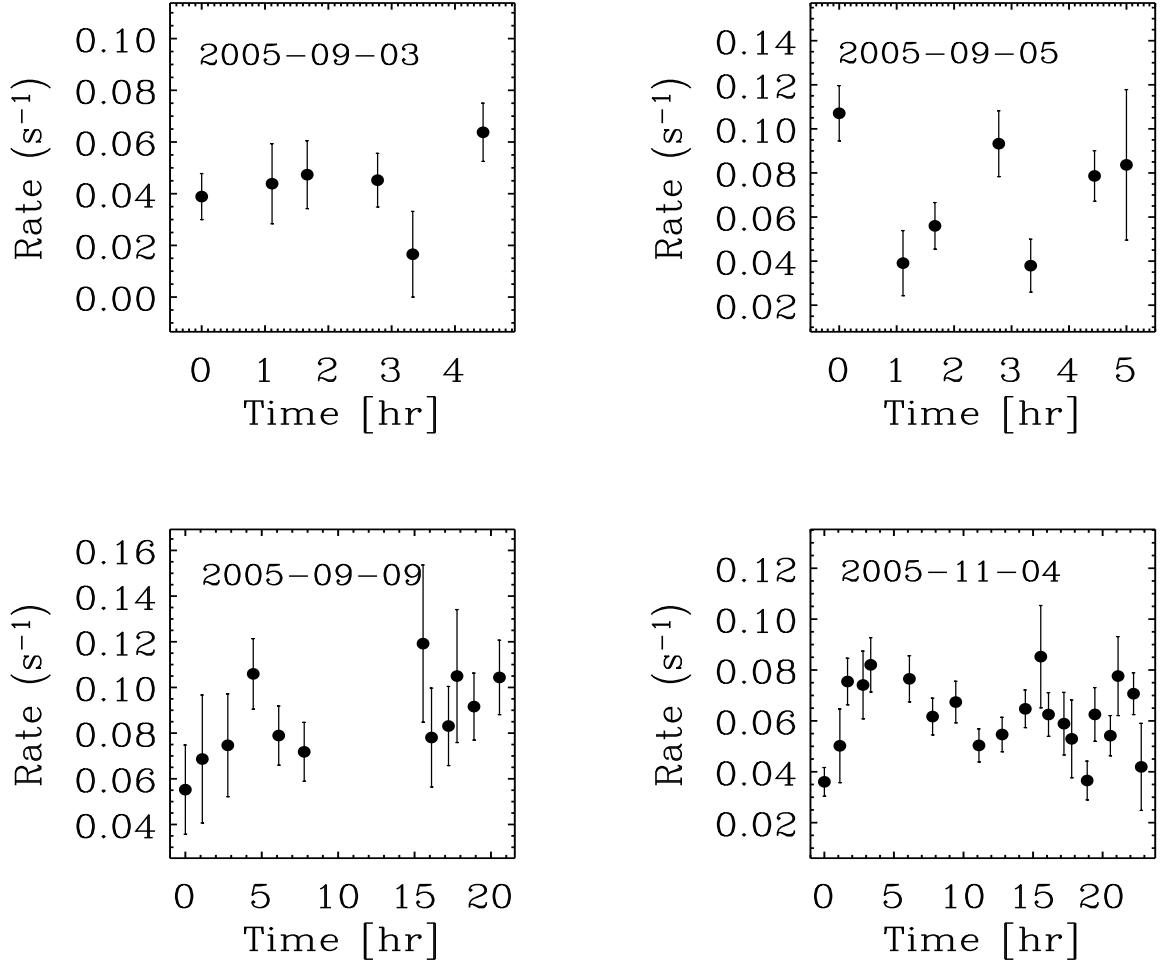,height=14.0cm,width=17.0cm}
}
\end{center}
\vspace{-1.0cm}
\caption{
{The \swift\ XRT light curves in total band (0.5--8~keV) at the four
epochs. Significant flux variability is present in the Nov 4 and Sept
5 observations. The flux changed by a factor 2.5 in one hour
in September and by  a factor
\gtsima 2 in 10 days in November. There is no accompanying spectral 
variability as indicated by the hardness ratio light curve which is
consistent with a constant (see text). 
}}

\end{figure}

\newpage 


\begin{figure}[h]
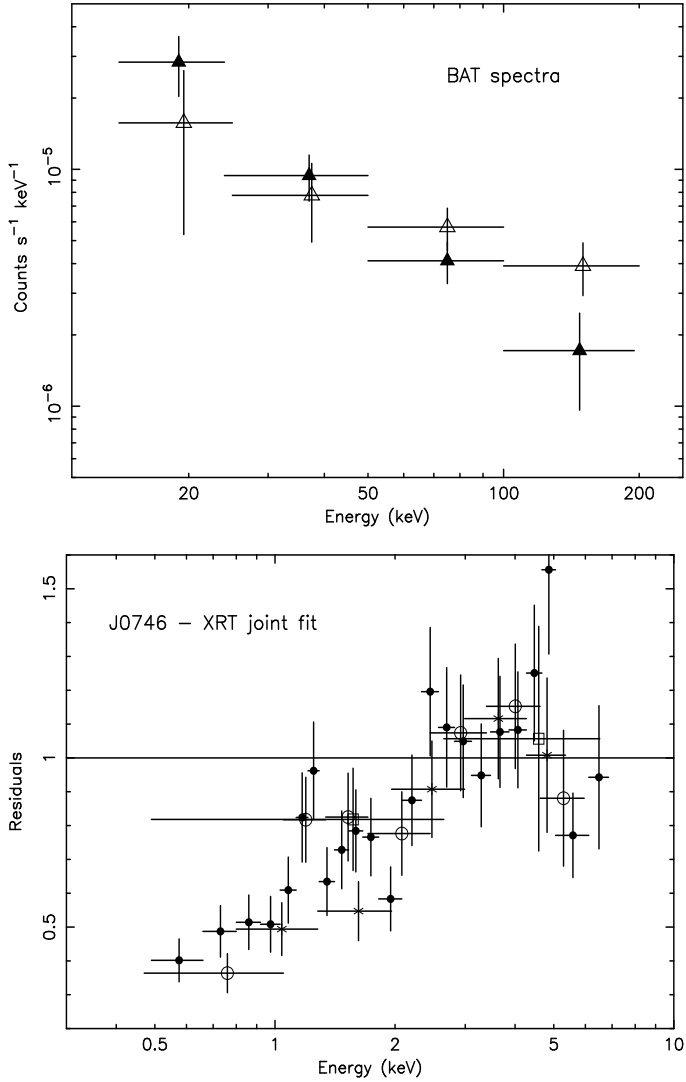

\begin{center}
\hbox{
\psfig{figure=f5a.ps,height=7.0cm,width=9.0cm,angle=-90}
}
\vspace{.3cm}
\hbox{
\psfig{figure=f5b.ps,height=7.0cm,width=9.0cm,angle=-90}
}
\end{center}
\vspace{-1.0cm}
\caption{
{\bf Top:} The BAT spectra from the 3 months {\it (open triangles)} and 9
months {\it (filled triangles)} integrations. There is a suggestion for the
3-months spectrum to be harder, albeit at only \gtsima 1$\sigma$. 
{\bf Bottom:} Residuals of a joint fit to the XRT spectra of J0746 with a
power law and Galactic absorption only. The data from the four epochs
of observations were fitted simultaneously in the energy range
2--8~keV, with the lower-energy datapoints added back. This plot
clearly illustrates the spectral flattening below 3 keV at all epochs.
{\it Open squares:} obsid 001; {\it Asterisks:} obsid 002; {\it Open circles:} obsid
003; {\it Filled squares:} obsid 007. }


\end{figure}

\newpage 


\begin{figure}[h]
\begin{center}
\hbox{
\psfig{figure=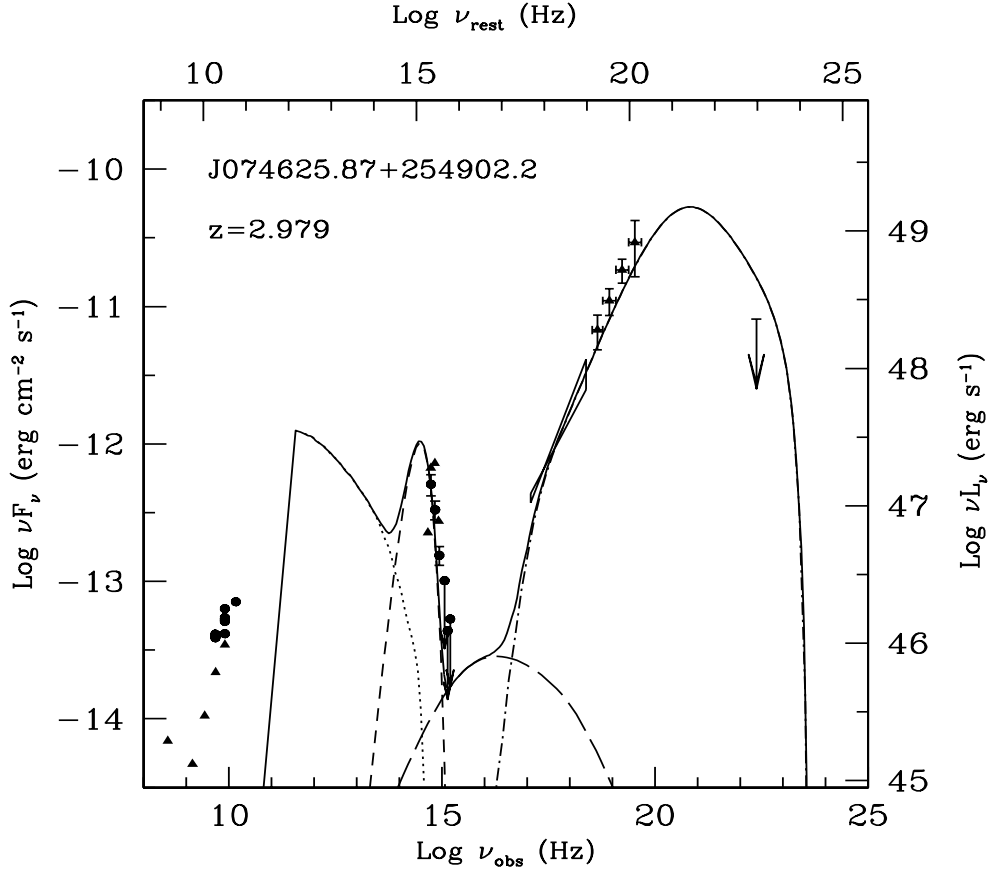,height=5.0in,width=5.5in}


}
\end{center}
\vspace{-1.0cm}
\caption{
\scriptsize
{Spectral Energy Distribution of J0746 from archival and coeval
observations. The BAT spectrum from the 9 months exposure is plotted,
together with the XRT spectrum, UVOT, and radio (filled circles)
fluxes from the November 4 campaign. The data plotted with triangles
are from NED, while the GeV upper limit is from our reanalysis of the
EGRET database. The SED is fitted with a composite jet+disk model. The
solid line shows the jet continuum calculated with the emission model
described in the text, assuming the following parameters:
$\Gamma=\delta=20$, $R=2.7\times 10^{16}$ cm, $B=2$ G, electron
density $n_e=9\times 10^4$ cm$^{-3}$, $\gamma _{\rm min}=1$, $\gamma
_{rm b}=50$, $\gamma_{\rm max}=10^3$, $n_1=1.5$, $n_2=3.8$. The
external radiation field has a luminosity $L_{\rm BLR}=10^{46}$ erg/s
and is diluted within the BLR assumed to have a radius
$R_{BLR}=10^{18}$ cm. The single emission components are also
reportes: synchrotron (dotted), SSC (long dashed), EC (dot-dash) and
the disk (dashed). The peak of the synchrotron component is produced
by the self-absorption.  }}

\end{figure}
\normalsize 
\newpage


\begin{figure}[h]
\begin{center}
\hbox{
\psfig{figure=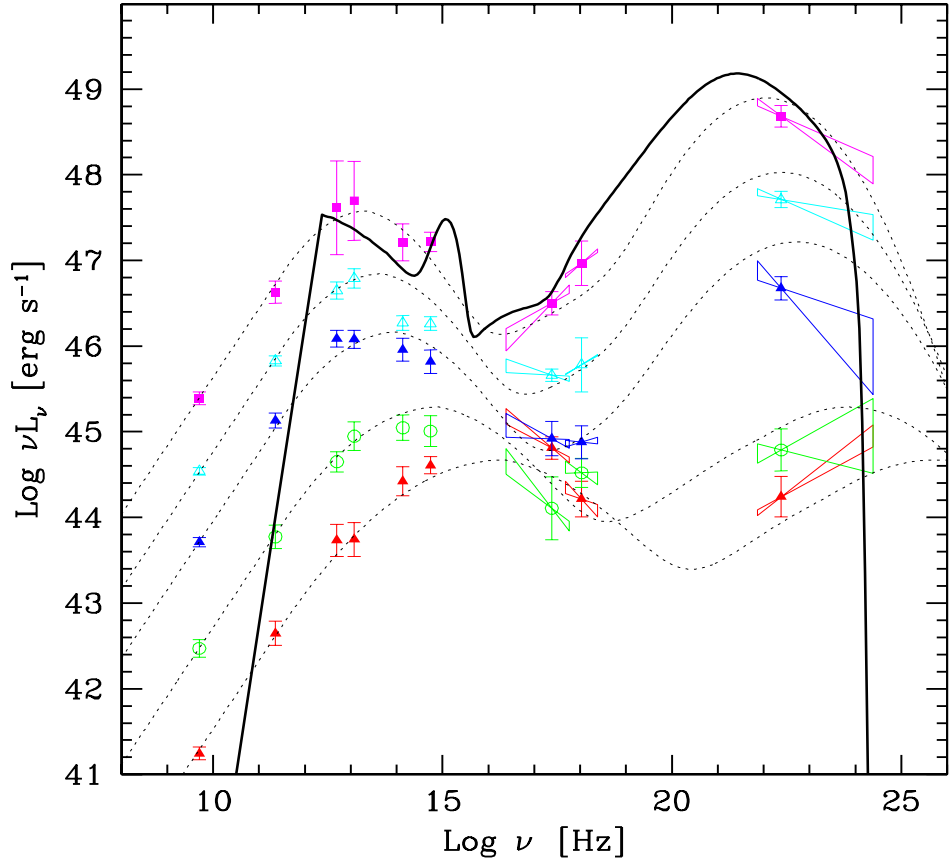,height=14.0cm,width=15.0cm}


}
\end{center}
\vspace{-1.0cm}
\caption{
{The Spectral Energy
Distributions (SEDs) of blazars form a sequence with luminosity
(Fossati et al. 1998). Going from luminous FSRQs to fainter TeV BL
Lacs, the synchrotron and IC peaks move forward, and the Compton flux
decreases. The solid black line is the best-fit model to the SED of
J0746, showcasing its extreme properties. 
}}

\end{figure}

\end{document}